\documentclass[aps,prl,showpacs,groupedaddress,twocolumn,preprintnumbers,amsmath,amssymb]{revtex4}

\usepackage{graphicx}% Include figure files
\usepackage{dcolumn}% Align table columns on decimal point
\usepackage{bm}% bold math
\usepackage{braket}

\begin{document}

\title{Quantum State Engineering using Single Nuclear Spin Qubit of Optically Manipulated Ytterbium Atom}% Force line breaks with \\

\author{Atsushi~Noguchi$^{1,2}$}
\author{Yujiro~Eto$^{2}$}
\author{Masahito~Ueda$^{2,3}$}
\author{Mikio~Kozuma$^{1,2}$}
\email[]{kozuma@ap.titech.ac.jp}
\affiliation{%
$^{1}$Department of Physics, Tokyo Institute of Technology, 2-12-1 O-okayama, Meguro-ku, Tokyo 152-8550, Japan}
\affiliation{%
$^{2}$ERATO Macroscopic Quantum Control Project, JST, 2-11-16 Yayoi, Bunkyo-Ku, Tokyo 113-8656, Japan}
\affiliation{%
$^{3}$Department of Physics, University of Tokyo, Hongo, Bunkyo-ku, Tokyo 113-0033, Japan}

\date{\today}% It is always \today, today,
             
\begin{abstract}
A single Yb atom is loaded into a high-finesse optical cavity with a moving lattice, and its nuclear spin state is manipulated using a nuclear magnetic resonance technique.
A highly reliable quantum state control with fidelity and purity greater than 0.98 and 0.96, respectively, is confirmed by the full quantum state tomography; a projective measurement with high speed $(500\ \mathrm{\mu s})$ and high efficiency $(0.98)$ is accomplished using the cavity QED technique. 
Because a hyperfine coupling is induced only when the projective measurement is operational, the long coherence times $\mathrm{(T_1\ =\ 0.49\ s\ and\ T_2\ =\ 0.10\ s)}$ are maintained. 
Our technique can be applied for implementing a scalable one-way quantum computation with a cluster state in an optical lattice.
% These values are sufficiently large to implement a scalable one-way quantum computation via the application of our technique to create a cluster state in an optical lattice.
\end{abstract}

\maketitle

Quantum computation enables us to solve NP problems such as factorization into prime factors and to perform quantum simulations of solid-state models\cite{1}. 
Although various research activities have been dedicated to the implementation of quantum computing, actual experiments have so far been limited to only few steps of quantum operations using a small number of qubits. 
Recently, an array of quantum gates was implemented for neutral atoms with massively parallel operation using optical lattice systems\cite{2,3,4}. 
However, in these experiments, the coherence time was limited to the order of milliseconds, which is too short to achieve scalable quantum computation\cite{5,6}. 
The other challenge is to address single sites of the lattice and control single qubits in a reliable manner.

It is advantageous to implement a qubit with a single nuclear spin\cite{7,8,9,10,11} because the nuclear spin is robust against a stray magnetic field; this robustness can be attributed to the weak strength of the magnetic moment generated by the nuclear spin as compared with that generated by the electronic spin\cite{12}. 
Here, we employ a single nuclear spin of the $^{171}\mathrm{Yb}$ atom in the form of a qubit in an optical lattice potential and implement both manipulation and state verification of the qubit with high fidelity and purity using nuclear magnetic resonance (NMR) and cavity QED techniques. 
It should be noted that a long coherence time of the qubit originates from an extremely weak coupling between the nuclear spin and the electromagnetic field and is normally traded off for enabling fast and highly efficient state measurement. 
We couple the electronic spin with the nuclear spin using hyperfine interaction in an excited state, and such a coupling is induced only when we irradiate the atoms with the probe light used for readout. 
We can thus maintain a long coherence time of the qubit. 
In the cluster computing, both high readout efficiency and high-speed detection are mandatory; these two conditions are satisfied in our experiment by using a cavity QED setup\cite{13,14}.
We will discuss the cluster quantum computation using the cavity QED system in the last section.
% cavity QEDを用いたクラスター量子計算の詳細については、論文の最後で議論する。
% Using any one of the techniques of atom sorting\cite{15}, optical accordion\cite{16}, or single-site addressing using ultranarrow transition\cite{9,10}, projective measurement of a single qubit in the cluster state can be performed using our cavity QED system.

%%%% Figure 1%%%%
\begin{figure}[b]
   \includegraphics[width=7cm]{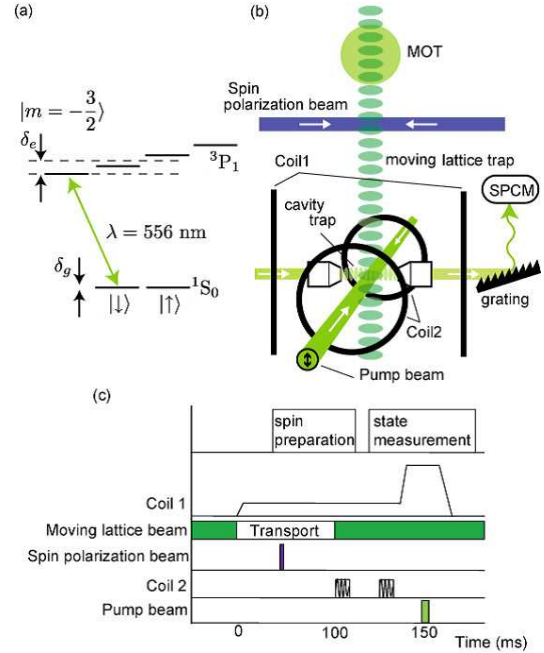}
\caption{Experimental setup and time chart. (a) Energy-level diagram of $^{171}$Yb. (b) Experiment setup. (c) Time chart for the experiment. $\ket{\uparrow}$ and $\ket{\downarrow}$ denote the magnetic sublevels $m_F= +1/2\ \mathrm{and} -1/2$ in the ground state $^1\mathrm{S_0 (F = 1/2)}$, respectively. The substates in the excited state $^3\mathrm{P}_1$ are labeled as $\ket{m_F}$.}
\label{setup}
\end{figure}
%%%%%%%%%%%%%
Our experimental setup, relevant energy levels, and the time chart of the experiment are shown in Fig. 1. 
We construct a qubit with a 1/2 nuclear spin of the ground state $\mathrm{^1S_0 (F = 1/2)}$ of the $^{171}\mathrm{Yb}$ atom and define the quantized axis along the cavity axis by using the bias magnetic field induced by ``Coil1." In the following discussion, $\ket{\uparrow}$ and $\ket{\downarrow}$  represent the magnetic sublevels of $\mathrm{^1S_0 (F = 1/2)}$. 
Details of Zeeman slowing and the double MOT system have been obtained from Refs. 17 and 18. 
To load the atoms into the cavity mode, we utilize the concept of a ``moving lattice." 
We first activate an optical standing wave potential, which is created by counter-propagating frequency-doubled $\mathrm{YVO}_4$ laser beams (532 nm), with a peak potential depth of $110\ \mathrm{\mu K}$ . 
Because the optical standing wave overlaps the MOT beams, atoms are loaded to the lattice potential. 
After the MOT beams and the relevant magnetic field are turned off, the lattice potential starts moving with a speed that is proportional to the value of $\delta$. 
By changing the value of $\delta$ according to the relation $\delta (t) = \delta _0\sin (\pi t/\tau)$ , atoms are transported to the cavity mode, where $\delta _0 = 2\pi \times 700\ \mathrm{kHz}$  and $\tau = 100\ \mathrm{ms}$ . 
To initialize the nuclear spin state to $\ket{\downarrow}$ , we irradiate the atoms in the moving lattice with a circularly polarized spin polarization beam that is resonant with the $\mathrm{^1S_0 (F = 1/2)\rightarrow}$ $\mathrm{ ^1P_1 (F' = 1/2)}$ transition (399 nm). 
When we manipulate the nuclear spin state, the radio frequency (RF) of magnetic field is applied to the atoms by using ``Coil2" after the atoms arrive in the cavity mode, where the radio frequency is resonant with the Zeeman splitting between two substates $\ket{\uparrow}$ and $\ket{\downarrow}$ (Fig. 1a, $\delta _g = 2\pi \times 2.5\ \mathrm{kHz}$ ). 
Because of the RF irradiation, the nuclear spins in the atoms rotate around the axis orthogonal to the cavity axis.
Our cavity consists of two concave mirrors having ultra-high reflectivity at 556 nm and is characterized by the following three parameters: the maximum interaction rate between atoms and photons $g_0 = 2\pi\times2.8\ \mathrm{MHz}$ , cavity decay rate (HWHM of the cavity resonance line) $\kappa = 2\pi\times 4.8\ \mathrm{MHz}$ , and atom decay rate (half natural linewidth) $\gamma = 2\pi\times 91\ \mathrm{kHz}$ . 
The length of our cavity is stabilized to $150\ \mathrm{\mu m}$  by injecting a 560-nm laser beam and utilizing the FM side-band method such that the cavity is resonant with the $\mathrm{^1S_0 (F = 1/2)\rightarrow}$ $\mathrm{ ^3P_1 (F' = 3/2)}$ transition (556 nm). 
It should be noted that the far-off resonant locking beam (560 nm) is responsible for trapping the atoms in the cavity mode, where the trap depth is  $30\ \mathrm{\mu K}$ (beam waist is $19\ \mathrm{\mu m}$ ). 
To perform the projective measurement, we increase the bias magnetic field and the Zeeman splitting in the excited state $\mathrm{^3P_1 (F' = 3/2)}$, where the splitting $\delta _e = 2\pi \times 60\ \mathrm{MHz}$ is much larger than the cavity-enhanced linewidth $\Gamma = \gamma\left[ 1+2g^2 /(\kappa \gamma)\right] = 2\pi\times 3.4\ \mathrm{MHz}$ for the $\mathrm{^1S_0 (F = 1/2)\rightarrow}$ $\mathrm{ ^3P_1 (F' = 3/2)}$ transition. 
When the atoms are irradiated with a linearly polarized pump beam resonant with the  $\mathrm{^1S_0 (F = 1/2)\rightarrow}$ $\mathrm{ ^3P_1 (F' = 3/2)}$ transition, only the $\ket{\downarrow}$ atoms are excited in a cyclic manner and repeatedly emit fluorescence photons into the cavity mode. Therefore, the detection of more than one photon implies that the nuclear spin is projected to $\ket{\downarrow}$ .
To perform experiments with single atoms, we prepare dilute atoms such that the expected mean atom number in the cavity mode becomes less than unity; utilize the post-selection method to perform the experiments. The total detection efficiency of a photon emitted from an atom is 0.1; this value is enhanced by the Purcell effect. We obtain four photons from each single atom on average. On the basis of the assumption of the binomial distribution for photodetection, the efficiency of the projective measurement is estimated to be 0.98. The result obtained from the projective measurement contain errors that can be attributed to the dark counts of the detector and also to unwanted spin flips caused by the excitation of the $\ket{\uparrow}$ state. These lead to about 2\% error for the diagonal elements of the density matrix.

%%%% Figure 2%%%%
\begin{figure}[t]
   \includegraphics[width=8.5cm]{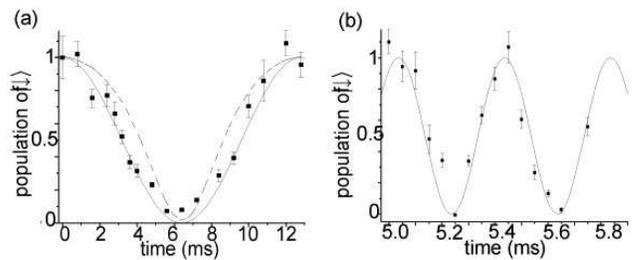}
\caption{Rabi oscillation and Ramsey interference. Time developments of population of $\ket{\downarrow}$ state as measured by Rabi oscillation (a) and Ramsey interference (b). The error bars are estimated on the basis of the statistical distribution of signal counts.}
\label{rabi}
\end{figure}
%%%%%%%%%%%%%
The time chart (Fig. 1c) of the experiment consists of two parts, i.e., the preparation and detection of the nuclear spin states. 
Arbitrary single nuclear spin states are prepared by rotating nuclear spins around two orthogonal axes. 
The rotation around the cavity axis is accomplished using the Larmor precession generated by the bias magnetic field, and the other rotation is implemented using Rabi oscillation through irradiation of the resonant RF field.
Figure 2a shows the Rabi oscillations of the single nuclear spin generated by the RF field. 
We induce Zeeman splitting between the two ground states $\ket{\uparrow}$ and $\ket{\downarrow}$ using the bias magnetic field and apply a square-shaped of the RF pulse whose frequency is resonant with the splitting $\delta = 2\pi\times 2.5\ \mathrm{kHz}$. 
The population of the down spin state as a function of the RF irradiation time is shown in Fig. 2a. Experimental results show that the visibility of 0.96, which is estimated by sinusoidal fitting. 
This high visibility enables us to precisely control the nuclear spin rotation around one axis. 
The fact that the operation is carried out using single nuclear spins can be confirmed as follows. 
In our projective measurement, we consider that the spin is projected to $\ket{\downarrow}$ when the photodetector counts more than one photon. 
Suppose that there are two atoms in the cavity mode. 
In this case, the photon counts are obtained even if the spin state is $\ket{\downarrow\downarrow}$ , $\ket{\downarrow\uparrow}$ , or $\ket{\uparrow\downarrow}$ , and thus, the variation in the probability of detecting more than one photon is no longer sinusoidal (see a dotted curve in Fig. 2a). 
The curve obtained from the experimental results represents a simple sinusoidal curve, which indicates the successful post-selection of a single nuclear spin.

To check the controllability of two-axis rotations, we measure the Ramsey interference for a single nuclear spin. 
On the basis of the measurements of the Rabi oscillations (Fig. 2a), we use an RF field with a duration of 3.2 ms as a $\pi /2$ pulse. 
We irradiate single nuclear spins with two $\pi /2$ RF pulses, and the nuclear spins rotate around the cavity axis during the interval between the two pulses. 
The population of the down spin state as a function of the time interval between the two RF pulses is shown in Fig. 2b. The curve obtained from the experimental results is in good agreement with the theoretical curve, and the visibility is calculated from the sinusoidal fitting to be 0.99, which is consistent with the value expected from the error generated in the projective measurement. 
It should be noted that the obtained visibility of Ramsey interference (0.99) is slightly higher than that obtained from the Rabi oscillations. 
In our experiment, the stability of the bias magnetic field is better than that of the RF field. We believe that the above difference arises because the RF irradiation time in the case of the Ramsey interference experiment is shorter than that in the case of the Rabi oscillation experiment. 
Hereafter, we use only $\pi /2$ pulses of the RF field. By appropriately varying the timing of the $\pi /2$ pulses, superposition states and also eigenstates are prepared.

%%%% Figure 3%%%%
\begin{figure}[t]
   \includegraphics[width=8.5cm]{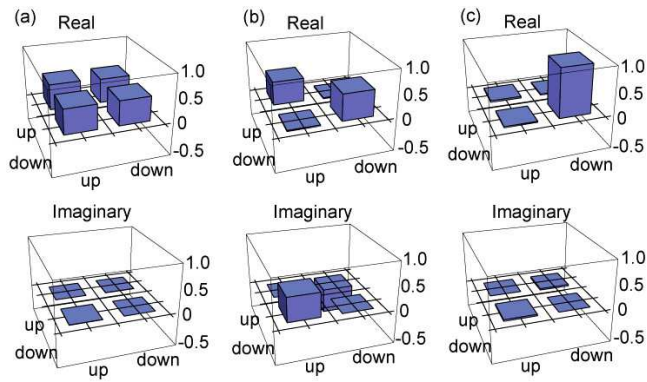}
\caption{Full quantum state tomography. Real and imaginary parts of density matrices for the prepared nuclear spin states reconstructed using the maximum likelihood estimation. The corresponding spin states are (a) $\left( \ket{\uparrow}+\ket{\downarrow}\right) /\sqrt{2}$ , (b) $\left( \ket{\uparrow}+i\ket{\downarrow}\right) /\sqrt{2}$ , and (c) $\ket{\downarrow}$.}
\label{tomo}
\end{figure}
%%%%%%%%%%%%%
By using spin polarization and rotation, we prepare arbitrary states of single nuclear spins.
Here three typical spin states are prepared, i.e., (a) $\left( \ket{\uparrow}+\ket{\downarrow}\right) /\sqrt{2}$ , (b) $\left( \ket{\uparrow}+i\ket{\downarrow}\right) /\sqrt{2}$ , and (c) $\ket{\downarrow}$ ; these sates are defined on a rotating frame with a frequency of  $\delta_g$ determined by the bias magnetic field. At t = 0 ms, we begin transporting single nuclear spins from the MOT to the cavity with the moving lattice. 
At t = 50 ms, the nuclear spin is polarized to the down state using the spin polarization beam (399 nm). Superposition states are created by applying the $\pi /2$ pulse at (a) t = 100.0 ms and (b) t = 100.1 ms. 
To reconstruct the density matrix ($\rho$) for each state, we perform projective measurements for the single nuclear spins along three directions. To perform the projective measurements along three orthogonal axes, we rotate the spins using Rabi oscillation or Lamor precession techniques before irradiating the atoms with the pump beam. 
Figure 3 shows the density matrix ($\rho$) for each state reconstructed by using the maximum likelihood estimation method. 
The purity ($p=\mathrm{Tr}\left[ \rho ^2\right]$) and the fidelity to the ideal state ($\mathrm{f}=\bra{\psi_i}\rho\ket{\psi_i}$; $\ket{\psi_i}$ is the ideal state) are estimated to be (a) (p,f) = $(0.98\pm 0.01,\ 0.99\pm 0.005)$ , (b) (p,f) = $(0.96\pm 0.01,\ 0.98\pm 0.005)$  and (c) (p,f) = $(0.97\pm 0.03,\ 0.98\pm 0.02)$ ; the Monte Carlo method is employed for this estimation assuming a binomial distribution for the signal counts. 
The purities and fidelities for the Fig. 3 are slightly less than 1, which are limited by the distortion of the RF field and photon scattering due to the moving lattice.
One-way quantum computing requires a highly efficient projective measurement because the computation is based on the result of individual projection and the scalability is therefore determined by the detection efficiency. 
Efficient and fast detection are realized in our experiment by using hyperfine interaction and also the enhanced mode-selective spontaneous emission due to the Purcell effect derived from the cavity QED system\cite{19}. 
Although the detection efficiency is determined by the rate of photon count (4 counts/500 $\mathrm{\mu s}$) in our experiment, a dramatic improvement (0.9998) can be expected by simply collecting atomic fluorescence from both sides of the cavity. Such an extremely high value of efficiency enables us to perform thousands of operations in the one-way quantum computing.

%%%% Figure 4%%%%
\begin{figure}[t]
   \includegraphics[width=7cm]{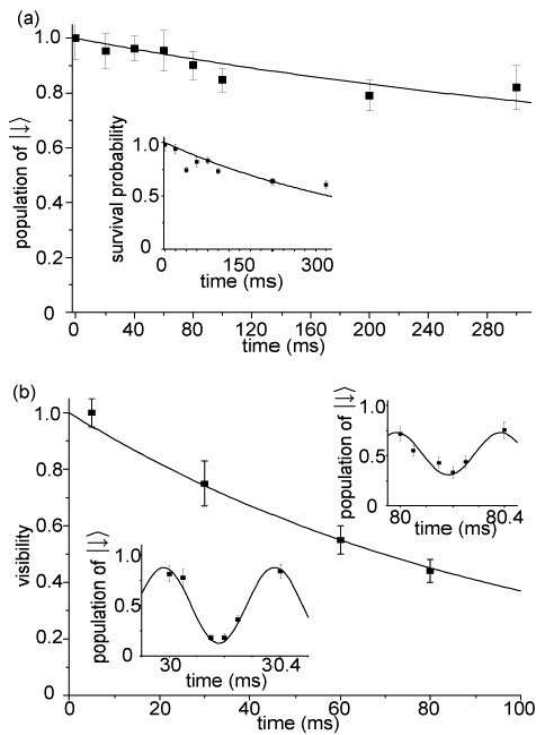}
\caption{Measurement of coherence time of single nuclear qubit. (a) Decay of population of $\ket{\downarrow}$ state. The inset shows the survival probability of the atom in the cavity as a function of time. (b) Time development of visibility of Ramsey fringe. The inset shows typical examples of Ramsey interference patterns. For both cases (a) and (b), the error bars are evaluated on the basis of the binomial distribution of signal counts.}
\label{tomo}
\end{figure}
%%%%%%%%%%%%%
To evaluate the longitudinal relaxation time $T_1$, we first measure the lifetime $\tau$ of single Yb atoms in the micro-cavity mode. 
We prepare the spin state of $\ket{\downarrow}$ and measure the population of $\ket{\downarrow}$ as a function of the trapping time. 
Next, the spin state of $\ket{\uparrow}$ is prepared, and the dependence of the population of $\ket{\downarrow}$ on the trapping time is measured. 
The sum of them gives the survival probability of single Yb atoms in the micro-cavity mode as a function of the trapping time (see the inset of Fig. 4a); the lifetime $\tau$ is estimated to be $0.44\pm 0.03\ \mathrm{s}$ by fitting the data to an exponentially decaying function. 
Figure 4a shows the population of $\ket{\downarrow}$ as a function of the trapping time; the fitting is obtained by normalizing the population of $\ket{\downarrow}$ with the survival probability. 
From the fitting, the longitudinal relaxation time $T_1$ is estimated to be $0.49\pm 0.15\ \mathrm{s}$. 
To estimate the transverse relaxation time $T_2$, we measure the visibility of the Ramsey interference as a function of the trapping time (Fig. 4b). 
The insets of Fig. 4b show the typical Ramsey interferences, where the horizontal axis represents the time separation between two RF pulses and the vertical axis represents the population of the down spin state. 
The transverse relaxation time $T_2$ is estimated to be $0.10\pm 0.01\ \mathrm{s}$ from the fitting. 
It should be noted that in the case of neutral alkali atoms, the coherence time of Zeeman sublevels is only of the order of $10\ \mathrm{\mu s}$\cite{7}. Even in the case in which both the clock states of different hyperfine substates and the spin echo technique are used, the coherence time is limited to 30 ms\cite{8}. The coherence time of 0.1 s obtained in our experiment is thus much longer than that of alkali atoms.
Here, we discuss the possible sources responsible for the decoherence. The experimentally obtained values of $T_1$ and $T_2$ are limited by photon scattering due to the moving lattice and by the fluctuation or inhomogeneity of the bias magnetic field. Because the fluctuation of the bias magnetic field does not affect the population of the spin state, $T_1$ is considered to be determined by the photon scattering rate $\Gamma_p$  according to the relation $T_1 = 1/\Gamma_p$ . The photon scattering rate $\Gamma_p$ is estimated to be $2.0\ \mathrm{s^{-1}}$ from the experimental parameters including the imperfection of mode match of counter propagating beams. The inverse of $\Gamma_p$ is 0.5 s which is in agreement with the experimentally obtained value of $T_1 = 0.49\ \mathrm{s}$.
Here, we define the decoherence rate $\Gamma_m$ , which originates from the energy-conserving dephasing effect due to the fluctuation and inhomogeneity of the bias magnetic field. Because transverse relaxation occurs due to longitudinal relaxation and the dephasing effect, the relation $1/T_2 = 1/T_1 +\Gamma_m$ should be satisfied. From the experimentally obtained values of $T_1$ and $T_2$, $\Gamma_m$ is estimated to be $8\ \mathrm{s^{-1}}$, which corresponds to the magnetic field fluctuation of 9 mG. The ratio of the coherence time and the time required for the projective measurement is 200; this ratio approximately gives the possible operation number. 
The suppression of the magnetic field fluctuation below 1 mG is technically possible; suppression to such values will increase the coherence time to values up to 0.5 s.

Our cavity QED system can be applied to implement a large scale of the cluster quantum computation. 
We prepare the nuclear spin qubit per site of an optical lattice using the fermionic band insulator state. 
To create the macroscopic cluster state for the nuclear spins, the nuclear spin dependent potential is required. 
While such a potential can be created by a laser beam whose detuning is less than hyperfine splitting, the ratio of the natural linewidth to the hyperfine splitting is not enough to suppress the spontaneous emission in the presence of ordinary dipolar allowed transitions of alkaline atoms. 
However, the problem can be solved by making an optical potential with the $\mathrm{^1S_0\rightarrow}$ $\mathrm{ ^3P_2}$ ultra-narrow transition of ytterbium. To create the cluster state, we use the Ising-type interaction based on the magnetic dipole moment of $\mathrm{ ^3P_2}$ state because the scattering length of the $\mathrm{^{171}Yb}$ atom is too small. 
Since the $\mathrm{ ^3P_2}$ state is automatically mixed when we induce the optical potential, the required interaction can be realized by simply making close the distance between neighboring atoms by the nuclear spin dependent potential. 
The required phase flip is estimated to take $10$ ms. 
Individual addressing is implemented with a magnetic gradient and a specific atom is excited to the $\mathrm{ ^3P_2}$ state using the Rabi oscillation. 
The nuclear spin state is transferred to the spin state of the Zeeman sub-levels of $\mathrm{ ^3P_2}$ state. The atom in the $\mathrm{ ^3P_2}$ state is transported to the micro cavity with our belt-conveyer technique and it is returned to the nuclear spin state in the micro cavity. 
We can perform the selection of basis and projective measurement on this qubit by using the method presented in this article. The cluster computation is thus accomplished by repeating these processes.

We would like to thank M. Takeuchi, N. Takei, P. Zhang, T. Mukaiyama, T. Kishimoto, and S. Inouye for their stimulating and fruitful discussions.

\end{document}